\documentclass[twocolumn, twocolappendix]{aastex701}
\usepackage{amssymb,amsmath,amsfonts,graphicx,epsf}
\usepackage{makecell}

\newcommand{\gtorder}{\mathrel{\raise.3ex\hbox{$>$}\mkern-14mu
            \lower0.6ex\hbox{$\sim$}}}
\newcommand{\ltorder}{\mathrel{\raise.3ex\hbox{$<$}\mkern-14mu
            \lower0.6ex\hbox{$\sim$}}}

\shorttitle{The Radius of PSR~J0437$-$4715}
\shortauthors{Miller, Dittmann, Holt, et al.}

\usepackage{hyperref}

\begin{document}

\title{THE RADIUS OF PSR J0437$-$4715 FROM NICER DATA
}

\correspondingauthor{M.~C.~Miller}
\email{mcmiller@umd.edu}

\author[0000-0002-2666-728X]{M.~C.~Miller}
\affiliation{Department of Astronomy and Joint Space-Science Institute, University of Maryland, College Park, MD 20742-2421 USA}
\email{miller@astro.umd.edu}

\author[0000-0001-6157-6722]{A.~J.~Dittmann}
\affiliation{Institute for Advanced Study, 1 Einstein Drive, Princeton, NJ 08540, USA}
\altaffiliation{NASA Einstein Fellow}
\email{dittmann@ias.edu}

\author[0000-0002-3097-942X]{I.~M.~Holt}
\affiliation{Department of Astronomy and Joint Space-Science Institute, University of Maryland, College Park, MD 20742-2421 USA}
\email{imholt@umd.edu}

\author[0000-0002-3862-7402]{F.~K.~Lamb}
\affiliation{Illinois Center for Advanced Studies of the Universe and Department of Physics, University of Illinois at Urbana-Champaign, 1110 West Green Street, Urbana, IL 61801-3080, USA}
\affiliation{Department of Astronomy, University of Illinois at Urbana-Champaign, 1002 West Green Street, Urbana, IL 61801-3074, USA}
\email{fkl@illinois.edu}

\author[0000-0003-2759-1368]{C.~Chirenti}
\affiliation{Department of Astronomy, University of Maryland, College Park, MD 20742-2421, USA}
\affiliation{Astroparticle Physics Laboratory NASA/GSFC, Greenbelt, MD 20771, USA}
\affiliation{Center for Research and Exploration in Space Science and Technology, NASA/GSFC, Greenbelt, MD 20771, USA}
\affiliation{Center for Mathematics, Computation, and Cognition, UFABC, Santo Andr{\'e}, SP 09210-170, Brazil}
\email{chirenti@umd.edu}

\author[0009-0008-6187-8753]{Z.~ Arzoumanian}
\affiliation{X-Ray Astrophysics Laboratory, NASA Goddard Space Flight Center, Code 662, Greenbelt, MD 20771, USA}
\email{zaven.arzoumanian-1@nasa.gov}

\author[0000-0003-4962-145X]{J.~Berteaud}
\affiliation{NASA Goddard Space Flight Center, Code 662, Greenbelt, MD 20771, USA}
\affiliation{Department of Astronomy, University of Maryland, College Park, MD 20742-2421, USA}
\email{berteaud@umd.edu}

\author[0000-0002-9870-2742]{S.~Bogdanov}
\affiliation{Columbia Astrophysics Laboratory, Columbia University, 550 West 120th Street, New York, NY 10027 USA}
\email{slavko@astro.columbia.edu}

\author[0000-0001-7115-2819]{K.~C.~Gendreau}
\affiliation{X-Ray Astrophysics Laboratory, NASA Goddard Space Flight Center, Code 662, Greenbelt, MD 20771, USA}
\email{keith.c.gendreau@nasa.gov}

\author[0000-0002-6089-6836]{W.~C.~G. Ho}
\affiliation{Department of Physics and Astronomy, Haverford College, 370 Lancaster Avenue, Haverford, PA 19041, USA}
\email{who@haverford.edu}

\author[0000-0003-4357-0575]{S.~M. Morsink}
\affiliation{Department of Physics, University of Alberta, Edmonton, AB, T6G 2E1, Canada}
\email{morsink@ualberta.ca}

\author[0000-0002-5297-5278]{P.~S. Ray}
\affiliation{Space Science Division, U.S. Naval Research Laboratory, Washington, DC 20375, USA}
\email{paul.s.ray3.civ@us.navy.mil}

\author[0000-0003-4815-0481]{R.~A.~Remillard}
\affiliation{MIT Kavli Institute for Astrophysics \& Space Research, MIT, 70 Vassar Street, Cambridge, MA 02139, USA}
\email{ronrem4@gmail.com}

\author[0000-0002-9249-0515]{Z. Wadiasingh}
\affiliation{Department of Astronomy, University of Maryland, College Park, MD 20742-2421, USA}
\affiliation{Astrophysics Science Division, NASA Goddard Space Flight Center, Greenbelt, MD 20771, USA}
\affiliation{Center for Research and Exploration in Space Science and Technology, NASA/GSFC, Greenbelt, MD 20771, USA}
\email{zorawar.wadiasingh@nasa.gov}

\author[0000-0002-4013-5650]{M.~T.~Wolff}
\affiliation{Space Science Division, U.S. Naval Research Laboratory, Washington, DC 20375, USA}
\email{MTWolff@cox.net}

\begin{abstract}

Neutron star Interior Composition Explorer (NICER) data have been used to estimate the masses and radii of the rotation-powered millisecond pulsars PSR~J0030$+$0451, PSR~J0740$+$6620, PSR~J0437$-$4715, PSR~J1231$-$1411, and PSR~J0614$-$3329, sometimes in joint analyses with X-ray Multi-Mirror (XMM-Newton) data. These measurements provide invaluable information about the properties of cold, catalyzed matter beyond nuclear saturation density.  Here we present the results of our modeling of NICER data on PSR~J0437$-$4715 using several different models of hot thermal X-ray emitting spots on the stellar surface.  For this pulsar, previous Nuclear Spectroscopic Telescope Array (NuSTAR) observations established that there is also a modulated nonthermal component to the emission, but the previously published analysis of NICER data did not model this component.  We find that the Bayesian evidence is significantly higher when the modulated nonthermal component is included, and that omission of this component leads to poor fits to the bolometric NICER data and thus risks bias in the resulting radius estimates.  Our models, which we pursue to inferential convergence, therefore have modulated nonthermal emission, and our headline model has in addition three uniform-temperature thermally-emitting circular spots.  Using this model, the symmetric 68\% credible range in the radius is 11.8~km to 15.1~km, which at the independently-measured mass of $M=1.418\pm 0.044~M_\odot$ is consistent with previous reports of the radius of the $\sim 1.4~M_\odot$ pulsar PSR~J0030$+$0451.  We discuss the implications of this measurement for the equation of state of dense matter.

\end{abstract}

\keywords{Neutron stars; Neutron star cores; Nuclear physics; X-ray astronomy}

\section{INTRODUCTION}
\label{sec:introduction}

The equilibrium state of cold matter beyond nuclear saturation density $\rho_{\rm sat}\approx 2.6\times 10^{14}~{\rm g~cm}^{-3}$ cannot be determined from first principles, because of the fermion sign problem (e.g., \citealt{1990PhRvB..41.9301L}; see \citealt{2019ARCMP..10..337L} for a recent review).  Even innovative frameworks such as chiral effective field theory \citep{1990PhLB..251..288W,1991NuPhB.363....3W,1992PhLB..295..114W,2021ARNPS..71..403D} have uncertainties that become large at densities greater than $\sim 1.5-2~\rho_{\rm sat}$.  Thus experiment and observation are essential to understand cold dense matter, and because terrestrial laboratories cannot reach the combination of high density, temperature much less than the Fermi temperature, and large excess of neutrons over protons seen in neutron stars, observations of neutron stars hold the key to understanding nuclear physics in this extreme state.

Over the last decade or so, several lines of astronomical observations have contributed constraints on the macroscopic properties of neutron stars, and thus on the microscopic equation of state (EOS; for cold, catalyzed neutron stars this is conveniently represented as the pressure $P$ as a function of energy density $\epsilon$).  These lines include the discovery of high-mass neutron stars by radio timing \citep{2010Natur.467.1081D,2013Sci...340..448A,2020NatAs...4...72C,2021ApJ...915L..12F,2025ApJ...983L..20S}, including the more model-dependent masses inferred from the so-called ``black widow" pulsars (e.g., \citealt{2022ApJ...934L..17R,2025arXiv251205099R}); the binary neutron star coalescence GW170817, which provided an upper limit to the tidal deformability of $\sim 1.4~M_\odot$ neutron stars \citep{2017PhRvL.119p1101A,2018PhRvL.121i1102D}; and X-ray measurements of the neutron stars PSR~J0030$+$0451 \citep{2019ApJ...887L..24M,2019ApJ...887L..21R,2024ApJ...961...62V}, PSR~J0740$+$6620 \citep{2021ApJ...918L..28M,2021ApJ...918L..27R,2024ApJ...974..295D,2024ApJ...974..294S}, PSR~J0437$-$4715 \citep{2024ApJ...971L..20C}, PSR~J1231$-$1411 \citep{2024ApJ...976...58S}, and PSR~J0614$-$3329 \citep{2025arXiv250614883M} using Neutron Star Interior Composition Explorer (NICER) and X-ray Multi-Mirror (XMM-Newton) X-ray data complemented by radio measurements.  To this list we can add potentially more model-dependent inferences from the electromagnetic afterglow of GW170817 and from general properties of short gamma-ray bursts \citep{2013PhRvL.111m1101B,2015ApJ...812...24F,2015ApJ...808..186L,2017ApJ...850L..19M,2018PhRvL.120z1103M,2018ApJ...852L..25R,2018PhRvD..97b1501R,2019ApJ...884L..16C,2023Natur.613..253C,2025ApJ...983...88G} and other X-ray data \citep{2016EPJA...52...63M,2016ApJ...820...28O,2016RvMP...88b1001W,2017A&A...608A..31N}.  By 2030 it may be possible to measure the moment of inertia of PSR~J0737$-$3039A via pulsar timing at radio wavelengths to a precision of $\sim 10$\% \citep{2020MNRAS.497.3118H}.

Here we report our analysis of NICER data on the 173.7~Hz pulsar PSR~J0437$-$5715.  Like the other pulsars observed with NICER, PSR~J0437$-$4715 is not actively accreting.  Instead, the X-ray pulsations evident in NICER data are caused by hot spots on the surface of the spinning neutron star, formed when highly relativistic (characteristic Lorentz factors likely $\gtorder 10^7$) beams of electrons and positrons, which are generated as part of the process that produces coherent radio emission, penetrate the surface and deposit their energy \citep{2011ApJ...743..181H}. PSR~J0437$-$4715 is in a binary system, so radio observations place tight constraints on the mass ($M=1.418\pm 0.044~M_\odot$), the distance ($D=156.96\pm 0.11$~pc, from the Shklovskii effect \citep{1970SvA....13..562S} on the orbital period derivative), and the inclination angle of the observer to the orbital axis ($i=137.506 \pm 0.016$~deg; all numbers from \citealt{2024ApJ...971L..18R}).  

Because PSR~J0437$-$4715 has the highest X-ray flux among non-accreting millisecond pulsars (with, e.g., $\sim 3\times$ the NICER count rate of PSR~J0030$+$0451 and $\sim 30\times$ the NICER count rate of PSR~J0740$+$6620), models of the X-ray hot spots are put to a particularly exacting test for this pulsar.  We therefore explore three models of the thermally emitting hot spots (all of which use nonmagnetic model atmospheres of fully ionized hydrogen) to the NICER data (see Section~\ref{sec:methods}).  Our models also include contributions which do not vary in a way commensurate with the pulsar rotational frequency and thus are unmodulated in the folded pulse waveform, based on a model of background counts from \citet{2022AJ....163..130R} and from an angularly nearby active galactic nucleus (AGN).  We further include a modulated nonthermal contribution, which we model as a power law, based on the observed modulation in higher-energy Nuclear Spectroscopic Telescope Array (NuSTAR) data \citep{2016MNRAS.463.2612G}.  

The observation of pulsed emission with NuSTAR in the energy range between $2-20$~keV \citep{2016MNRAS.463.2612G} is significant because it is implausible that received flux with this spectrum and energy range could be produced by thermal emission originating on the surface of the neutron star. Furthermore, \citet{2016MNRAS.463.2612G} showed that a power law emission model describes both the NuSTAR data and the lower energy (0.5--2.0 keV) XMM-Newton observations \citep{2013ApJ...762...96B}  of a nonthermal spectral component. This implies that a small fraction of the pulsed emission appearing in the NICER data (0.3--3.0 keV) originates from a region co-rotating with the neutron star, such as the magnetosphere. We model this emission with a modulated power law component, as described in Section \ref{sec:nonthermal}. As shown in Section \ref{sec:models}, our models that include this modulated power law result in much better Bayesian evidence than the models that attempt to describe all of the NICER emission as arising from thermally emitting hot spots on the star's surface.

Our spot models are two uniform-temperature circular spots, two uniform-temperature oval spots, and three uniform-temperature circular spots.  The spots have independent angular sizes and temperatures, can be anywhere on the star, and can overlap or not, as the fit prefers. As we discuss in Section~\ref{sec:fits-to-data}, when the modulated power law is not included, the bolometric fit to the data is poor.  When the modulated power law is included, the bolometric fit is better and all three of the models with a modulated power law have similar Bayesian log evidence, which is several tens larger than when the modulated power law is not included.  Thus the modulated power law must be included, but none of the spot models is preferred on the basis of their Bayesian evidence.  The three-circle model has the highest log likelihood; its radius posterior is the broadest of the three models (although the radius posterior overlaps at the 90\% credible level for all models).  We therefore feature this model. Its symmetric 68\% credible range for the radius is 11.8~km to 15.1~km.  

We have also compared our best NICER-only model with the X-ray Multi-Mirror (XMM-Newton) data on PSR~J0437$-$4715.  We find that extra XMM-Newton background flux, beyond that measured in the immediate vicinity of the pulsar, is necessary to obtain agreement between our model and the XMM-Newton data.  We therefore focus entirely on fits of the NICER data, rather than attempting joint fits with the XMM-Newton data.

In Section~\ref{sec:observations} we describe our selection of data and the calibration of the instruments.  In Section~\ref{sec:methods} we give a brief description of our analysis methods; more details may be found in previous papers \citep{2019ApJ...887L..24M,2021ApJ...918L..28M,2024ApJ...974..295D}.  In Section~\ref{sec:models} we discuss our models and compare our best fit with the data, finding no systematics in the residuals.  In Section~\ref{sec:fits-to-data} we present our results and in Section~\ref{sec:EOS} we discuss the implications that our PSR~J0437$-$4715 radius measurements, combined with other results, have for the equation of state of cold, catalyzed matter above $\rho_{\rm sat}$.  In Section~\ref{sec:conclusions} we present our conclusions, and in several appendices we provide details of our fits including corner plots. Posterior samples from our analysis will be made available upon request to the corresponding author.

\section{OBSERVATIONS}
\label{sec:observations}

For this analysis we make use of NICER X-ray Timing Instrument (XTI) exposures of PSR J0437--4715 covering the interval between 2017 July 6 and 2021 July 31. The NICER data of this pulsar contain a significant flux contribution from the bright Seyfert
II AGN RX J0437.4$-$4711 \citep{1996ApJ...464..760H}, situated $4\farcm18$ away. To minimize contamination due to the AGN and  other neighboring sources,  NICER observations of PSR J0437--4715 were carried out using a pointing position $1\farcm47$ to the southwest of the pulsar \citep[see][for further details]{2019ApJ...887L..25B}.

To produce a clean event list suitable for analysis, we adopt the 3C50 model approach, primarily because it is designed to also produce a background spectrum from the source data set with reliable uncertainty estimates in the energy band of interest. The details of the 3C50 model are described in \citet{2022AJ....163..130R}.

Valid exposure segments for PSR J0437$-$4715 were first determined with the HEASoft tool \texttt{nimaketime}, and the times were masked by $\pm 30$ s at the orbital day/night boundary. From there, continuous intervals longer than 200\,s were selected for further cleaning. For each good time interval (GTI), the raw NICER spectrum is extracted, a background prediction is obtained with the 3C50 model, and six filtering tests (see the next paragraph) are performed on the results.
Only GTIs with 50 detectors active are included in the final selection, with events from detectors with \texttt{DET\_ID} 14 and 34 always excluded as they frequently exhibit elevated count rates well above the average of the other detectors.

The 3C50 background model makes use of three non-source count rates from the on-source GTIs of the actual observations being considered to select and rescale background components from a reference library of models. Two filters apply to parameters needed for the background model: the slow chain noise rate normalized to 50 Focal Plane Modules (FPMs) at 0.0--0.25 keV, $nz < 220$ counts s$^{-1}$, and the normalized rate of good events at 15--18 keV (where the performance of the XTI is such that effectively no
astrophysical source signal is expected), $ibg < 0.2$ counts s$^{-1}$ ($nz$, $ibg$, and $hbg$ are defined in \citealt{2022AJ....163..130R}). The next two filters act on the net background-subtracted spectrum, $S0_{\rm net} < 0.15$ counts s$^{-1}$ and $hbg_{\rm net} < 0.05$ counts s$^{-1}$, where the corresponding energy bands are 0.2--0.3 keV for $S0$ and 13--15 keV for $hbg$. Finally, when the net pulsar spectrum shows very faint or undetectable intensity above 2\,keV, filters are added in the C band (2--4 keV), $C_{\rm net} < 0.1$ counts s$^{-1}$, and D band (4--12 keV), $D_{\rm net} < 0.3$ counts s$^{-1}$.
This rigorous filtering procedure reduces the initial unfiltered 2.736 Ms exposure to 1.310 Ms.

NICER also conducted on-axis observations of the AGN RX J0437.4$-$4711 during 2017 July 10--20, 2017 October 8--10, 2021 December 10--26, and 2022 February 28--March 19, in order to establish its flux and range of variability. 
The mean spectrum of the AGN was determined for each epoch using the 3C50 background model and level 2 filtering applied to each GTI. The average intensity (0.3--3.0 keV) was determined to be 6.26 counts s$^{-1}$, while the rms deviation between the four epochs is 25\%. The expected contribution from the AGN in the PSR J0437--4715 dataset is represented by the mean AGN spectrum, scaled down by a factor of 42, corresponding to the reduction in telescope effective area based on the angular separation between the AGN and the offset pointing direction.

The final selection of filtered NICER events was phase-folded based on the ephemeris from a publicly available Parkes Pulsar Timing Array radio timing solution \citep[see][]{2024ApJ...971L..18R} using the PINT\footnote{\url{https://github.com/nanograv/PINT}} \citep{2021ApJ...911...45L} \texttt{photonphase} tool. 

\section{METHODS}
\label{sec:methods}

Our approach to the generation of waveforms and the analysis of NICER and XMM-Newton data has been described in several previous papers.  We summarize our methods, with references to papers with more details, in the following subsections.

\subsection{Spacetime}

To generate waveforms we select heated patterns on a neutron star surface and trace those photons through vacuum to the observer using the ``oblate Schwarzschild approximation" (\citealt{2007ApJ...663.1244M,2014ApJ...791...78A}; for an update see \citealt{2025ApJ...994..163J}.  In this approximation, we model the surface as an oblate spheroid with properties given by the rotation rate and the mass and radius the star would have without rotation, and include all special relativistic effects.  However, we use the spherical Schwarzschild spacetime for the external spacetime of the star.  For rotation rates below $\sim 600$~Hz, the error introduced by this approximation, compared with the numerically exact solution obtained by solving the full Einstein field equations for a rigidly rotating star, is several times smaller than the statistical uncertainties in the data \citep{2019ApJ...887L..26B}.  We have also verified the accuracy of our waveform codes against other codes \citep{2019ApJ...887L..26B,2021ApJ...914L..15B,2024ApJ...975..202C}.

\subsection{Magnetic Fields}

We assume that the magnetic field in the neutron star atmosphere is too weak to affect the atmospheric structure or the spectrum or beaming of photons from the surface, at NICER energies.  In practice, this is a good approximation when the electron cyclotron energy is much less than the 0.3~keV minimum energies that we analyze, i.e., when the surface magnetic field is much less than $\sim 3\times 10^{10}$~G (for more detail, see \citealt{2021ApJ...914L..15B}).  PSR~J0437$-$4715 has a rotation period of $P=0.005757$ seconds and an observed period derivative ${\dot P}=5.729\times 10^{-20}$ \citep{2008ApJ...679..675V,2024ApJ...971L..18R}.  However, the majority of this observed period derivative comes from proper motion (the Shklovskii effect, \citealt{1970SvA....13..562S}), and when the measured proper motion etc. from \citet{2024ApJ...971L..18R} are used to calculate the magnitude of this effect we find that the intrinsic period derivative is only ${\dot P}_{\rm intrinsic}=1.369\times 10^{-20}$.  Inserting this into equation (12) of \citet{2006ApJ...643.1139C} implies that for a magnetic field inclination $\theta$ of $\pi/2$ and a field configuration with $\alpha=0$ (i.e., so that the closed region of the magnetic field reaches the light cylinder), the surface field strength is $B\approx 4\times 10^8$~G, which is weak enough to be neglected unless significantly stronger multipoles exist on the surface. See Section~3.1 of \citet{2021ApJ...918L..28M} and Section~3.1.3 of \citet{2024ApJ...974..295D} for more details about this argument.

\subsection{Atmosphere Models}
Because the neutron star surface gravity is so strong and the star is not actively accreting, we expect that the atmosphere will consist purely of the lightest element present \citep{1980ApJ...235..534A}.  Given the 5.74 day orbital period of the PSR~J0437$-$4715 binary \citep{2008ApJ...679..675V}, which is easily large enough to have contained a donor with a hydrogen envelope, the atmosphere is likely to be pure hydrogen. We attempted one run using a helium atmosphere, and found that the resulting fits were poor, with a low probability of being correct.  Therefore, our results focus on fully ionized nonmagnetic hydrogen atmospheres, using the NSX code \citep{2001MNRAS.327.1081H} for the model atmospheres.\footnote{Partial ionization is possible for the lower-temperature spots, but previous work has found little difference in the inferred radius between fully and partially ionized hydrogen atmospheres (see, e.g., section 3.1 of \citealt{2021ApJ...918L..28M}) and the tables for partially ionized hydrogen are not as complete as for fully ionized hydrogen.}

We work in the deep-heating approximation: that the energy emerging from the photosphere was effectively deposited infinitely deep in the atmosphere by the high-energy electrons and positrons that hit the surface as part of the pulsar process.  This is a good approximation at the Lorentz factors $\gamma\sim 10^7$ expected for millisecond pulsars \citep{2002ApJ...568..862H}.  If, contrary to expectations, most of the deposited energy comes from far lower-energy electrons and positrons, $\gamma<100$, then the energy is deposited in shallower layers.  In that case, \citet{2019ApJ...872..162B} and \citet{2020A&A...641A..15S} found, as expected, that the resulting beaming pattern of emission would be broader than in the deep-heating approximation.  This would lead to a lower modulation fraction in the waveform.  To match an observed modulation fraction, other parameters would therefore need to adjust.  In particular, a lower compactness $GM/(Rc^2)$ (for a star of gravitational mass $M$ and circumferential radius $R$) leads to a higher modulation fraction with all else equal, so if the energy is deposited in a shallow layer but the data are analyzed assuming deep heating one would expect that the inferred radius would be smaller than the actual radius.  

\citet{2025ApJ...982..112Z} found the opposite (that the inferred radii would be too large if the beaming pattern of emission is broader than assumed in analyses), but given that in their synthetic data they used (a)~$60\times$ as many spot counts as in the PSR~J0740$+$6620 data, with no background, (b)~$3\times$ the spot effective temperature that was inferred for PSR~J0740$+$6620, and (c)~a restricted spot distribution (two identical antipodal spots; NICER pulsars do not have antipodal spots) it is not clear that this result is applicable to the stars studied using NICER.

\subsection{Surface Emission Patterns}
Our hot spots on the stellar surface are constructed with either uniform-temperature circular spots or, in a generalization of that model, uniform-temperature oval spots.  The spots can overlap in an arbitrary way; we number our spots, and assume that a pixel in more than one spot emits with the effective temperature of the lowest-numbered spot that includes the pixel.  We do not expect that real heated regions on neutron stars come in these simple patterns, but previous work \citep{2009ApJ...705L..36L,2009ApJ...706..417L,2013ApJ...776...19L,2015ApJ...808...31M,2025arXiv251116759H} suggests that even when models deviate from the true pattern, if there is a statistically good fit then the inferred radius is not significantly biased.  This is understood to be because the angularly broad emission from a thermally emitting point on the surface, combined with gravitational light deflection, blur details of the spot patterns.

\subsection{Non-Pulsar NICER counts}
In addition to the NICER counts produced by photons from the spots, there will in general be counts produced in a number of other ways, including from particle background, optical loading from the Sun, other X-ray sources in the field whether resolved or unresolved, or unassociated X-ray emission from the pulsar system (such as from a pulsar wind nebula or from intrabinary shocks).  What these other counts have in common is that none of them are expected to be modulated at a frequency commensurate with the rotational frequency of the pulsar.  Our most general and conservative treatment of these ``background" counts in NICER data therefore assumes that in each NICER pulse invariant (PI) channel, independently, there is an additional component that is independent of rotational phase, and in our Bayesian analysis we marginalize over that component (see Section~3.4 of \citealt{2019ApJ...887L..24M} for more details).  In some cases we may have reliable information about one or more components of the background.  In the case of PSR~J0437$-$4715, the 3C50 data selection procedure discussed in Section~\ref{sec:observations} also outputs a background estimate, which we implement as discussed in Section~B.1 of \citet{2022ApJ...941..150S}.  There is also an AGN in the field of view, which we incorporate in our background estimates.

\subsection{The New Component: Modulated Non-thermal Pulsar Emission }
\label{sec:nonthermal}
An additional element that must be added in our analysis of PSR~J0437$-$4715 is the modulated nonthermal component, discovered in NuSTAR data by \citet{2016MNRAS.463.2612G}.  Because this component is modulated at the pulsar rotational frequency, it cannot be incorporated into the unmodulated background.  For simplicity, \citet{2024ApJ...971L..20C} did not include this component, but we show that when it is not included, the bolometric fit is poor and the inferred radius could be biased.  We have found that a good representation of the modulation is a Gaussian profile in the rotational phase.  We therefore model the photon number flux from the power law, at an energy $E$ and rotational phase $\psi$ using the expression 
\begin{equation}
\begin{array}{rl}
F_{\rm PL}(E,\psi)&=N_{\rm PL}(E/1~{\rm keV})^{-\alpha_{\rm PL}}\bigl[1+\\
&A_{\rm PL}\exp(-(\psi-\psi_{\rm PL})^2/(2(\Delta\psi_{\rm PL})^2))\bigr].\\
\end{array}
\end{equation}
This model describes the modulated nonthermal component using five parameters: the overall normalization $N_{\rm PL}$, the power law index $\alpha_{\rm PL}$, the fractional modulation $A_{\rm PL}$, the rotational phase $\psi_{\rm PL}$ of maximum modulation, and the width $\Delta\psi_{\rm PL}$ of the Gaussian.  The priors for the power law parameters, motivated by the NuSTAR observations, are listed in Table \ref{tab:wf-primary-parameters}.

\subsection{Statistical Methods}
Our overall statistical inference approach is Bayesian: we explicitly specify the models and their priors, use Bayes' Theorem to obtain the posteriors, and marginalize over nuisance parameters (such as background parameters) to obtain the posteriors for parameters of interest such as the mass and radius.  Because the parameter space is complex, with 20 to 30 parameters in the runs we discuss in Section~\ref{sec:fits-to-data}, it is necessary to use sophisticated statistical samplers to obtain posteriors.  In the next subsection, we discuss our sampling procedure in detail.  See Section~4 of \citet{2019ApJ...887L..24M} for tests of our sampling with synthetic two-circle and two-oval data similar to the NICER data on PSR~J0030$+$0451.

\subsection{Bayesian Methodology}
\label{sec:Bayesian}

We take a two-stage approach to evaluating models for the NICER data on PSR J0437$-$4715, similar to our analyses of the J0030$+$0451 and J0740$+$6620 data \citep{2019ApJ...887L..24M,2021ApJ...918L..28M,2024ApJ...974..295D}: in the first stage, we sample each model in a relatively agnostic manner, exploring the entire prior volume and estimating the Bayesian evidence for each model; in the second stage, we confirm or correct the results of the first stage using Markov chain Monte Carlo (MCMC) methods, which are less suited to efficient global exploration but have more reliable convergence properties. In the first stage we employed the \texttt{pocomc} sampler \citep{2022MNRAS.516.1644K}, which combines sequential Monte Carlo \citep[see, e.g.,][for a review]{2019arXiv190304797N} with normalizing flows \citep[e.g.,][]{JMLR:v22:19-1028} to achieve relatively robust and efficient sampling and evidence estimation without prior knowledge about which subdomains of the prior volume ought to be sampled. We follow up the initial \texttt{pocomc} analyses using the \texttt{emcee} sampler \citep{2013PASP..125..306F} to confirm or correct the inferred posterior distributions before we use them in EOS inference.  We find that the posterior after resampling with \texttt{emcee} is essentially indistinguishable from the posterior after a converged \texttt{pocomc} run.

\subsubsection{Sequential Monte Carlo Analyses}
\label{sec:poco}
We begin our analyses using the publicly available \texttt{pocomc} package,\footnote{\url{https://github.com/minaskar/pocomc}} which utilizes normalizing flows to precondition a given target distribution before sampling it using an adaptive sequential Monte Carlo scheme \citep{2022MNRAS.516.1644K}. For our purposes, the crucial benefits of this scheme are that a) it is able to estimate the Bayesian evidence for model comparison, b) it samples from the entire prior volume without relying on potentially biasing initialization schemes, and c) when applied to our NICER inferences it tends to outperform many packages that implement variations on nested sampling, which offer the aforementioned benefits \citep{2004AIPC..735..395S}. Specifically, the nested sampling algorithm MultiNest \citep{2009MNRAS.398.1601F} has been used extensively in previous NICER analyses \citep[e.g.,][]{2019ApJ...887L..24M,2019ApJ...887L..21R,2021ApJ...918L..28M,2021ApJ...918L..27R,2024ApJ...974..295D,2024ApJ...974..294S,2024ApJ...971L..20C}. Although MultiNest has a tendency to produce biased posterior and evidence estimates \citep[see][]{2016S&C....26..383B,2020AJ....159...73N,2023StSur..17..169B,2023MNRAS.521.1184L,2024OJAp....7E..79D}, it has provided useful starting points for more robust MCMC analyses \citep{2019ApJ...887L..24M,2021ApJ...918L..28M,2024ApJ...974..295D}. However the additional model complexity necessary to describe both the thermal emission from the surface of PSR J0437-4715 and the modulated power-law emission, presumably from its magnetosphere, proved too expensive for the methods used previously.\footnote{Notably, MultiNest is based on rejection sampling, so the cost of analyses grows exponentially with the number of model parameters. Algorithms based on Monte Carlo techniques have better (polynomial) asymptotic scaling. Nested sampling packages that can employ slice sampling, which has better scaling, include PolyChord \citep{2015MNRAS.450L..61H}, dynesty \citep{2020MNRAS.493.3132S}, and UltraNest \citep{2021JOSS....6.3001B}, but we have found \texttt{pocomc} \citep{2022MNRAS.516.1644K} better suited to our analyses.}

Sequential Monte Carlo \citep[e.g.,][]{10.1117/12.138224,doi:10.1049/ip-f-2.1993.0015,05168831-63ee-32a1-a34a-9a292ef12230, 2019arXiv190304797N} lies at the heart of \texttt{pocomc}, which is itself a generalization of importance sampling. Importance sampling by itself uses a known importance sampling density ($\rho(\theta)$, where $\theta$ represents the parameters) to guide the estimation of an unknown target density $p(\theta)$, which can expedite inference if the importance sampling density is similar to the target density, but can otherwise degrade performance severely: one cannot generally construct a useful importance sampling density a priori. Sequential Monte Carlo constructs a sequence of sampling and target densities, in this application sampling first from the prior $\pi(\theta)$ to estimate an annealed posterior $p_i(\theta)\propto\pi(\theta){\cal L}(\theta)^{\beta_i}$, where ${\cal L}(\theta)$ is the likelihood, for a sequence of $\beta_i$ ranging from 0 and 1, sampling from the prior to the posterior. By gradually adjusting $\beta_i$, the importance sampling density at each stage can be smoothly adapted so that the next stage functions properly. For this additional effort, one gains an estimate of the Bayesian evidence and the ability to sample efficiently from the posterior once the corresponding importance sampling density has been determined, while avoiding any need to make assumptions at the outset about which subsets of the prior to sample. 

MCMC methods can struggle with correlated or skewed target distributions.  This can be ameliorated by preconditioning, a process by which a complicated target distribution is transformed to a distribution that is simpler and more amenable to sampling. \texttt{pocomc} uses normalizing flows \citep[e.g.,][]{JMLR:v22:19-1028} to accomplish this task, which are generative models that, over the course of sampling, infer bijective mappings between a given distribution and a Gaussian distribution. Preconditioning accelerates the sequential Monte Carlo algorithm by roughly an order of magnitude when sampling nontrivial target distributions \citep{2022MNRAS.516.1644K}. 

While modeling PSR J0437$-$4715, we found that \texttt{pocomc} provides significantly more robust results than, for example, MultiNest analyses of comparable computational cost. Still, given the complicated high-dimensional likelihood surfaces encountered in our analyses, our \texttt{pocomc} analyses demanded significant convergence testing. The thoroughness of a given \texttt{pocomc} analysis is controlled by $N_{\rm effective}$, the effective number of weighted particles used to explore parameter space at each stage of the algorithm. While the default value is set to $N_{\rm effective}=2^9=512$, we found that values of $N_{\rm effective}\gtorder 2^{17}$ were necessary to thoroughly sample our most complicated models, in the sense that doubling the resolution no longer significantly affected the inferred log evidence values and the widths of various posterior distributions. We emphasize, however, that compared to other (often more expensive) sampling methods, \texttt{pocomc} can occasionally underestimate posterior widths \citep[e.g.,][]{2023MNRAS.525.3181L,2025arXiv250618977W}, although not to the same extent as MultiNest (e.g., \citealt{2019ApJ...887L..24M,2021AJ....162..237I,2021ApJ...918L..28M,2024ApJ...974..295D,2024ApJ...974..294S,2025arXiv251116759H}; see \citealt{2020MNRAS.497.3118H} for a recent comparison between MultiNest and UltraNest). Thus, for our models that include the necessary power-law components, we have followed up each \texttt{pocomc} analysis with additional MCMC sampling to ensure the accuracy of our reported posteriors.  We now discuss this procedure in more detail.

\subsubsection{Markov Chain Monte Carlo Analyses}\label{sec:mcmc}
Following each of our \texttt{pocomc} analyses, we used the estimated posterior distributions to draw initial walker positions for an MCMC analysis using the \texttt{emcee} package \citep{2013PASP..125..306F}, employing the affine-invariant ``stretch'' proposal of \citet{2010CAMCS...5...65G}. Because we utilized a proposal distribution that satisfied detailed balance, the distribution of walkers will eventually converge to sample from the target distribution directly. In practice, because the results produced by \texttt{pocomc} were very nearly converged, we typically observed that the tails of each posterior broadened slightly, by a few percent, after which we continued drawing samples to reduce discretization errors in our posterior estimates. 

Each analysis used $2^{13}=8192$ walkers, for which we drew initial positions by resampling a Gaussian kernel density estimate of the corresponding \texttt{pocomc} posterior distributions. We judged the convergence of each analysis by monitoring the 1st, 16th, 50th, 84th, and 99th percentiles of each distribution and identifying the point at which each stopped exhibiting secular variations in time. We terminated each analysis after accruing $\sim10^6$ (effectively) independent samples for each model.

\section{Models used in our analysis}
\label{sec:models}

\begin{deluxetable*}{c|l|c}
\setlength{\tabcolsep}{19pt} 
    \tablecaption{Primary parameters of the pulse waveform models considered in this work.}
\tablehead{
      \colhead{Parameter} & \colhead{Definition} & \colhead{Assumed Prior}
    \label{tab:wf-primary-parameters}
}
\startdata
      \hline
      $c^2R_e/(GM)$ & Inverse of stellar compactness & $3.2-8.0$ \\
      \hline
      $M/M_\odot$ & Gravitational mass & ${\cal N}(1.418,0.044)$\\
      \hline
      $\theta_{\rm c1}$ (rad) & Colatitude of spot 1 center & $0$ to $\pi$ \\
      \hline
      $\Delta\theta_1$ (rad) & Spot 1 half-extension& $0-3$ \\
      \hline
      $kT_{\rm eff,1}$ (keV) & Spot 1 effective temperature & $0.011-0.5$\\
      \hline
      $f_1$ & Spot 1 elongation factor &\makecell{ $\log_{10}f_1$ flat from $0$ to $+1$} \\
      \hline
      $\psi_1$ (rad) & Spot 1 tilt angle & $0-\pi$\\
      \hline
      $\Delta\phi_2$ (cycles) & Spot 2 longitude difference & $0-1$ \\
      \hline
      $\theta_{\rm c2}$ (rad) & Colatitude of spot 2 center & $0$ to $\pi$\\
      \hline
      $\Delta\theta_2$ (rad) & Spot 2 half-extension& $0-3$ \\
      \hline
      $kT_{\rm eff,2}$ (keV) & Spot 2 effective temperature & $0.011-0.5$ \\
      \hline
      $f_2$ & Spot 2 elongation factor & \makecell{$\log_{10}f_2$ flat from $0$ to $+1$} \\
      \hline
      $\psi_2$ (rad) & Spot 2 tilt angle & $0-\pi$ \\
      \hline
      $\Delta\phi_3$ (cycles) & Spot 3 longitude difference & $0-1$ \\
      \hline
      $\theta_{\rm c3}$ (rad) & Colatitude of spot 3 center & $0$ to $\pi$ \\
      \hline
      $\Delta\theta_3$ (rad) & Spot 3 half-extension & $0-3$ \\
      \hline
      $kT_{\rm eff,3}$ (keV) & Spot 3 effective temperature & $0.011-0.5$ \\
      \hline
      $f_3$ & Spot 3 elongation factor & \makecell{$\log_{10}f_3$ flat  from $0$ to $+1$} \\
      \hline
      $\psi_3$ (rad) & Spot 3 tilt angle & $0-\pi$  \\
      \hline
      $f_{\rm back,3C50}$ & Multiplier of 3C50  background spectrum & ${\cal N}(1,0.0461)$ \\
      \hline
      $f_{\rm back,AGN}$ & Multiplier of AGN  background spectrum & $0.5-1.5$\\
      \hline 
      $N_{\rm PL}$ (${\rm keV}^{-1}~{\rm cm}^{-2}~{\rm s}^{-1}$)& Power law normalization & \makecell{$\log_{10}N_{\rm PL}$ flat  from $-7.0$ to $-2.0$ }\\
      \hline
      $\alpha_{\rm PL}$ & Power law photon index & ${\cal N}(1.5,0.25)$ \\
      \hline
      $A_{\rm PL}$ & Power law modulation fraction & $0-1$\\
      \hline
      $\psi_{\rm PL}$ (cycles) & Power law phase & $0-1$ \\
      \hline
      $\Delta\psi_{\rm PL}$ (cycles)& Power law width & $0.01-0.49$ \\
      \hline
      $\theta_{\rm obs}$ (rad) & Observer inclination & ${\cal N}(0.74166,0.000279)$ \\
      \hline
      $N_H$ (cm$^{-2})$ & Neutral H column density & $(0-20)\times 10^{20}$ \\
      \hline
      $d$ (kpc) & Distance & ${\cal N}(0.15696,0.00011)$ \\
      \hline
      $f_{\rm eff}$ & NICER effective area factor & ${\cal N}(1,0.1)$
\enddata
\tablecomments{These are the parameters and priors for our most complex model (Model 6). This model has three uniform-temperature oval spots with arbitrary overlap, plus a modulated power law with a Gaussian phase profile to take into account the modulated nonthermal component visible in NuSTAR data on PSR~J0437$-$4715, plus prior information on the NICER background from the model of \citet{2022AJ....163..130R} and from the nearby AGN. Simpler models (e.g., with two oval spots or two or three circular spots) have the same priors for the parameters they have in common with the three-oval model.  Except where noted, the prior is flat over the given range.  ${\cal N}(a,b)$ means a Gaussian with mean $a$ and standard deviation $b$.}
\end{deluxetable*}

We employ several models in our analysis of the NICER data on PSR~J0437$-$4715.  Table~\ref{tab:wf-primary-parameters} has a description of the primary parameters for our most complex model (in terms of number of parameters), which has three uniform-temperature oval spots as well as a modulated power law to represent the modulated nonthermal component. Less complicated models (with two circular spots, two oval spots, or three circular spots) have the same priors as the three-oval model for the parameters they have in common.

In our fits we incorporate information about known components of the unmodulated NICER background (see \citealt{2022ApJ...941..150S} for similar use of NICER background information in the analysis of PSR~J0740$+$6620).  In the case of PSR~J0437$-$4715 this includes not just the ``3C50" NICER instrumental background model of \citet{2022AJ....163..130R}, but also counts from the angularly nearby AGN RX J0437.4$-$4711 \citep{1996ApJ...464..760H}. For both sources of information, we produced a smoothed count spectrum using a cubic spline.  We then added to the phase-channel model produced from the spots a background that was an energy-independent factor times the nominal background count spectrum; the priors on the factors are listed in Table~\ref{tab:wf-primary-parameters}.  Because of our assumption that the background counts are not modulated at a frequency commensurate with the rotation frequency of PSR~J0437$-$4715, we distribute these extra counts uniformly in rotational phase.

In addition to these known sources of backgrounds, we allow for unknown additional sources of background in the NICER data.  As described in more detail in Section~3.4 of \citet{2019ApJ...887L..24M}, we do this independently in each NICER PI channel, and marginalize over the added background by fitting a Gaussian to the likelihood near the peak and performing an analytic integration of the marginalization integral.  Thus in most of our fits there is a lower limit to the total background counts in each NICER PI channel based on known backgrounds, but because we use a Gaussian prior on the 3C50 contribution this is not a hard limit.

In more detail, the models we fit are as follows.  In each case, the emission from all spots uses a fully ionized, nonmagnetic hydrogen atmosphere, and we fit only to the NICER data rather than performing a joint NICER$-$XMM fit.  Moreover, in all models but one the spots can be anywhere on the star, with or without overlaps. All of our models include background contributions from the 3C50 model of \citet{2022AJ....163..130R} as well as the angularly nearby AGN.  Our six models are summarized in Table~\ref{tab:model-list} and described in more details below:

\begin{enumerate}

\item Two circular spots with no modulated power law.

\item Three circular spots with no modulated power law.

\item Two circular spots with a modulated power law.

\item Two oval spots with a modulated power law.

\item Three circular spots with a modulated power law. Model 5 is our featured model.

\item Three oval spots with a modulated power law. This model did not converge.

\end{enumerate}

Models 1 -- 5 appeared to converge, based on a comparison of independent \texttt{pocomc} runs with different values of the precision parameters.  However, at the next level of complexity (Model 6 with three oval spots with a modulated power law), we were not able to establish convergence.  For example, at the highest resolution that we were able to employ in the three-oval run, the maximum log likelihood was $\sim 8$ less than the maximum log likelihood for the three-circle run, even though three circles is a subset of three ovals.  We therefore stopped our model development at the Model 5.

\begin{deluxetable}{c|l|c|c}
    \tablecaption{ 
    Description of the pulse waveform models considered in this work.}
\tablehead{
      \colhead{\makecell{Model \\number}} & \colhead{Spot model} & \colhead{Modulated PL?} & \colhead{Converged?}
    \label{tab:model-list}
}
\startdata
      \hline
      1 & Two circles & No & Yes\\
      \hline 
      2 & Three circles & No & Yes \\
      \hline
      3 & Two circles & Yes & Yes\\
      \hline
      4 & Two ovals & Yes & Yes\\
      \hline
      \textbf{5} & \textbf{Three circles} & \textbf{Yes} & \textbf{Yes} \\
      \hline
      6 & Three ovals & Yes & No \\
\enddata
\tablecomments{In every model  we incorporate estimates of the NICER background from the empirical 3C50 model \citep{2022AJ....163..130R} and the angularly nearby AGN, and there is no maximum set on the background.  Moreover, all models use fully ionized, deep-heating, nonmagnetic hydrogen for the atmosphere, and in each case we analyze only the NICER data rather than performing a joint NICER$-$XMM analysis. All models assume uniform-temperature spots. The bold font highlights our featured model.}
\end{deluxetable}

\section{RESULTS OF THE ANALYSIS OF THE \textit{NICER} PULSE WAVEFORM DATA}
\label{sec:fits-to-data}

\subsection{Quality of fits and inferred radii}

Table~\ref{tab:model-results} summarizes the qualities of the best fit of each of the models that we consider, along with the $\pm 1\sigma$ range of the inferred radius in each model.  We see that all of the models have phase-channel $\chi^2$ probabilities high enough that the fit could be acceptable.  However, we also see that the maximum log likelihood and log evidence are much worse for the models (1 and 2) that do not have a modulated power law component.  The remaining three models overlap in their $\pm 1\sigma$ radius intervals.  Based on the quality of the bolometric fits, we will use only the three-circle posterior when we discuss the implications of our results for the equation of state of cold, catalyzed matter beyond nuclear saturation density.  

\begin{deluxetable*}{c|l|l|l|l|c}
    \tablecaption{Summary of the results of the analysis using our pulse waveform models. The maximum likelihoods and evidences in the second and third columns are relative to our featured model, i.e., our model number 5, highlighted in bold font.}
\tablewidth{0pt}
\tablehead{
      \colhead{Model number} & \colhead{$\Delta\ln{\cal L}_{\rm max}$} & \colhead{$\Delta\ln{\cal Z}$} & \colhead{Phase-channel $\chi^2$/dof (prob)} & \colhead{Bolometric $\chi^2$/dof (prob)} & \colhead{$\pm 1\sigma$ $R_{\rm eq}$ (km)}
    \label{tab:model-results}
}
\startdata
      \hline
      1 &  $-48.93$ & $-25.31$ & 8359.14/8347 (0.460) & 36.13/27 (0.112) & $12.11-13.76$ \\
      \hline 
      2 & $-38.69$ & $-36.56$ & 8371.1/8352 (0.427) & 47.97/27 (0.008) & $10.25-12.66$ \\
      \hline
      3 &  $-8.76$ & $-0.40$ & 8352.72/8349 (0.49) & 38.92/27 (0.0644) & $13.60-15.96$ \\   
      \hline
      4 & $-8.15$ & $-1.48$ & 8350.76/8345 (0.48) & 39.64/27 (0.055) & $13.94-16.10$\\
      \hline 
      \textbf{5} & $0$ & $0$ & \textbf{8335.00/8345 (0.53)} & \textbf{33.39/27 (0.18)} & \textbf{11.82~--~15.14}\\
      \hline
\enddata
\tablecomments{The effective number of parameters which affect the bolometric light curve is much smaller than the total number of parameters, and is mostly independent of the model; here we estimate that there are effectively 5 parameters that affect the bolometric light curve.}
\end{deluxetable*}

\begin{figure}
          \includegraphics[width=1.0\linewidth]{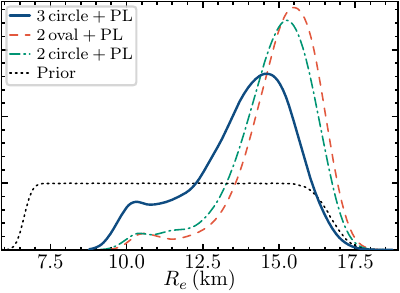}
\caption{Radius posteriors for each of our three models with a modulated power law (the vertical axis is linear in the probability density); from top to bottom in the legend the model numbers are 5, 4 and 3.  The dotted line shows the prior.  There is a large overlap of the posteriors of all three models, which shows that among the models with a modulated power law, the posterior is not especially sensitive to the  model chosen.}     
\label{fig:Rcred}
\end{figure}

Figure~\ref{fig:Rcred} compares the radius posteriors for our converged models with a modulated power law, i.e., models 3, 4 and 5.  Although the posteriors are not identical, they have substantial overlap.  This demonstrates that, at least when a modulated power law is included, the posterior is not sensitive to the details of the spot model.

See the Appendix for the full posteriors for our fit with three circular spots and a modulated power law (model number 5). 

\subsection{Characteristics of the best-fit model with three circular spots and a modulated power law}

\begin{figure}

          \includegraphics[width=\linewidth]{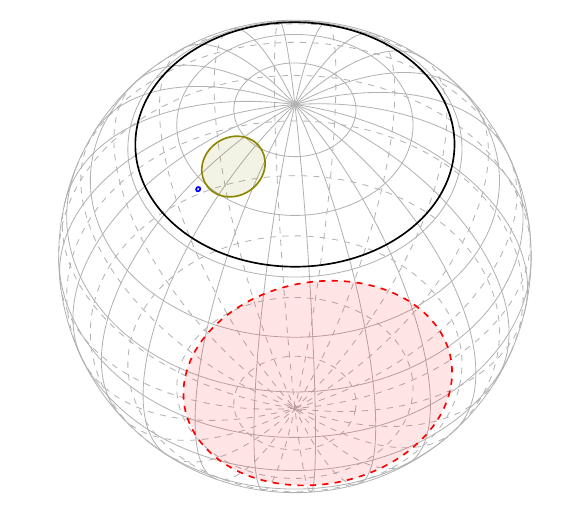}

\caption{Spot locations, sizes, and temperatures for the best-fit of our featured model with three uniform-temperature circular spots plus a modulated power law, i.e., model number 5.  The smallest spot has an effective temperature, as measured by a comoving observer on the surface, of 0.17~keV; the middle-sized spot has an effective temperature of 0.031~keV; and the largest spot has an effective temperature of 0.11~keV.  The solid black circle indicates the colatitude of the observer, 0.742~radians, which is strongly constrained by radio observations.  See the Appendix for the full set of parameter values for this best fit.  Bearing in mind that the temperature distribution is surely not actually uniform circles, this could be an indication that the two spots in the hemisphere of the observer represent a single spot with a range of temperatures.}     
\label{fig:spots}
\end{figure}

In this section we explore the characteristics of the best fit we obtained to the NICER PSR~J0437$-$4715 data using our featured model, i.e., model number 5, which has three circular spots plus a modulated power law with a Gaussian phase profile.  

Figure~\ref{fig:spots} shows the spot distribution in the best fit.  Two of the spots are close to each other and in the same hemisphere as the observer, albeit with very different temperatures, whereas the third spot, which has a much larger solid angle, is in the other hemisphere.  This could be a configuration that approximates one spot with a spatially asymmetric distribution of temperatures, and one with a more uniform temperature distribution.

\begin{figure}

          \includegraphics[width=\linewidth]{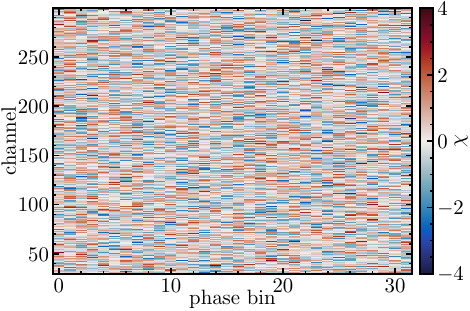}

\caption{Phase-channel residuals for the best fit to the data of our model with three uniform-temperature circular spots plus a modulated power law, over the full set of 32 uniformly-spaced rotational phases and NICER PI channels 30--299 inclusive.  Here, for a phase-channel bin $i$, if the data have $d_i$ counts and the model predicts $m_i$ counts, we define $\chi\equiv (m_i-d_i)/\sqrt{m_i}$.  In addition to the fit having an overall acceptable $\chi^2/{\rm dof}=8335.00/8345$ (for which the probability of this $\chi^2$ or higher for a correct model is 53\%), there are no patterns visible and no individual phase-channel bins with an unexpectedly large $|\chi|$.  This is a one-way test: our satisfactory fit does not guarantee that the model is correct, but a very low probability would indicate that we would need to look more closely at our model.}     
\label{fig:residuals}
\end{figure}

Figure~\ref{fig:residuals} shows the residuals, which we compute using
\begin{equation}
    \chi={\rm (data-model)}/({\rm model})^{1/2}
\label{eq:chi}
\end{equation}
for the best fit and the data for each NICER PI channel (vertical axis) and the rotational phase (horizontal axis).  No patterns are visible, and no individual phase-channel bin has a significantly larger $|\chi|$ than would be expected with this number of bins.  This, in addition to the overall $\chi^2/{\rm dof}=8335/8345$ (probability $0.53$ if the model is correct), shows that this test does not raise any concerns.  

\begin{figure}
          \includegraphics[width=\linewidth]{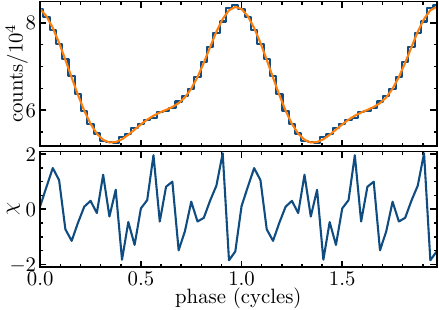}
\caption{(top panel) Comparison between the bolometric data (histogram) and the best model with three uniform-temperature circular spots plus a modulated power law (red line).  (bottom panel) Residuals between the data and the best model.  Here we plot two full rotational cycles, with 32 uniformly-distributed rotational phases per cycle.  In addition to the overall adequate bolometric $\chi^2/{\rm dof}=33.39/27$ (18\% probability if the model is correct), we see no strong outliers and no obvious patterns in the residuals.  As with the phase-channel $\chi$, this is a one-way test, which might detect a strong deviation between the model and the data but cannot guarantee the model's correctness if it passes the test.
}     
\label{fig:boloresiduals}
\end{figure}

Figure~\ref{fig:boloresiduals} shows the bolometric best-fit model and data (upper panel) and residuals (again using Equation~\ref{eq:chi}), as a function of rotational phase.  As with the phase-channel $\chi$, in addition to the fit being formally acceptable ($\chi^2/{\rm dof}=33.39/27$, which has a probability of 0.18 if the model is correct), there are no evident patterns or strong outliers and thus this test does not suggest any problems with the fit.  The bolometric and phase-channel comparisons between the best fit and the data yield one-way tests: if the fit were poor we would need to examine the models closely, but the fact that the fits are acceptable does not guarantee that the model is correct.

\begin{figure}
          \includegraphics[width=\linewidth]{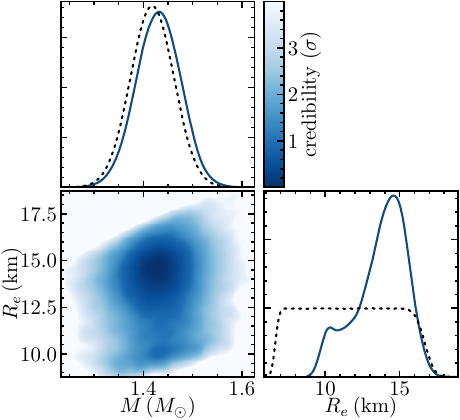}

\caption{Mass and radius posteriors, using our best fit to the data of our model with three uniform-temperature circular spots plus a modulated power law, i.e., our model number 5.  The dotted lines show the priors.  The approximate upper diagonal boundary in the mass-radius plot corresponds to our $c^2R_{\rm e}/(GM)=8$ prior upper limit, whereas the approximate lower boundary is set by the likelihood rather than by our prior $c^2R_{\rm e}/(GM)=3.2$ lower limit.  The mass posterior is single-peaked and is shifted only slightly from the tight prior given by radio observations, whereas the radius posterior is bimodal (with a larger mode at higher radii).}     
\label{fig:mr}
\end{figure}

Figure~\ref{fig:mr} shows the mass-radius posterior for our fit using three circular spots plus a modulated power law, i.e., our model 5.  The dotted curve on the one-dimensional mass plot (top left) shows the prior from the radio observations of \citet{2024ApJ...971L..18R}, from which it is clear that the radio observations provide most of the information about the mass.  The bottom right panel shows the one-dimensional radius posterior and demonstrates that it is bimodal with one mode predominating.  

\subsection{Comparison with Choudhury et al.\ (2024)}

We now compare these results with the earlier analysis of NICER observations of PSR J0437$-$4715 by \citet{2024ApJ...971L..20C}, which analyzed the same data set that we have analyzed, using the X-PSI code \citep{2023zndo...7632629R}, and did not include a modulated power law  component. More specifically, we performed runs using the ST (single-temperature) + PDT (protruding double temperature) configuration  of \citet{2024ApJ...971L..20C} (first introduced in \citealt{2019ApJ...887L..21R}).  ST+PDT is a restricted subset of a full three-circle configuration in which (a)~none of the three spots are allowed to have angular radii greater than $\pi/2$, (b)~spot 1 is not allowed to overlap with spot 2 or spot 3, and (c)~spots 2 and 3 must overlap, at least at a point.  For this configuration and with no constraints on the background (labeled "No BKG" in Figure 6 of \citealt{2024ApJ...971L..20C}), \citet{2024ApJ...971L..20C} obtained a $\pm 1\sigma$ radius range of 9.81--10.46~km.  This is not the headline model of \citet{2024ApJ...971L..20C}, but the handling of the background is closer to one of our options, and thus this provides a convenient point of comparison.

 The radius range we find for the no-background ST+PDT configuration with no modulated power law, using parallel-tempered \texttt{emcee} (see Section 3.5 of \citet{2021ApJ...918L..28M} for methodological details), is $10.16-12.21$~km at $\pm 1\sigma$, which is much broader than reported by \citet{2024ApJ...971L..20C}.  In previous parameter estimations for other pulsars, such as PSR J0740$+$6620 \citep{2024ApJ...974..295D,2024ApJ...974..294S}, our radius posteriors have typically been wider than those produced by X-PSI analyses, which is probably due to differences in the statistical sampling in the two methods. 
 Our best fit for our version of the ST+PDT model has an acceptable phase-channel $\chi^2/{\rm dof}=8375.25/8352$ (probability of 0.427), but the bolometric $\chi^2/{\rm dof}=47.97/27$ has a probability of only 0.008 and therefore might indicate a poor fit. 

A comparison of our results, with and without a modulated power law, and the results of \citet{2024ApJ...971L..20C} (which do not include a modulated power law) demonstrates that in addition to being required for a good bolometric fit the power law component shifts the posterior to higher radii.  Thus if the modulated power law is not included it risks bias in the radius posterior.

\section{IMPLICATIONS FOR THE EOS}
\label{sec:EOS}

\subsection{Statistical Method}
\label{sec:statistical}

As in \citet{2019ApJ...887L..24M}, \citet{2021ApJ...918L..28M}, and \citet{2024ApJ...974..295D}, we base our EOS inference on the method described in \citet{2020ApJ...888...12M}.  This is a fully Bayesian approach in which we begin with a selection of EOSs (three families of which are described below), which by fiat all have the same prior probability.  The (unnormalized) posterior probability of each EOS is then updated by calculating the product of the likelihoods of several data sets: these include the high masses of PSR~J1614$-$2230 ($M=1.937\pm 0.014~M_\odot$; \citealt{2023ApJ...951L...9A}) and PSR~J0348$+$0432 ($M=1.806\pm 0.037~M_\odot$; \citealt{2025ApJ...983L..20S}); the tidal deformability of two double neutron star coalescences seen in ground-based gravitational-wave detectors \citep{2017PhRvL.119p1101A,2020ApJ...892L...3A}; the mass and radius measurements of PSR~J0030$+$0451 \citep{2019ApJ...887L..24M} and PSR~J0740$+$6620 (\citealt{2021ApJ...918L..28M}); and now our mass and radius measurement of PSR~J0437$-$4715.  At densities up to half of nuclear saturation (i.e., up to roughly the crust-core transition density; see \citealt{2013ApJ...773...11H}) we assume that the QHC19 \citep{2019ApJ...885...42B} EOS applies, but because uncertainties at densities this low do not contribute significantly to uncertainties in the neutron star mass and radius, other choices of the low-density EOS give very similar results.  

We perform our likelihood calculations for maximum mass, radius as a function of mass, and tidal deformability as a function of mass by assuming that the star is rotating much slower than its maximum possible rate (typically $\sim 1500$--2000~Hz for most EOSs at $1.4~M_\odot$) and thus that the Tolman-Oppenheimer-Volkoff equations (TOV; \citealt{1939PhRv...55..364T,1939PhRv...55..374O}) are accurate.  They are highly accurate for the 173.7~Hz rotation rate of PSR~J0437$-$4715, as can be seen in Figure~13 of \citet{2019ApJ...887L..24M}, which demonstrates that the difference between the equatorial radius of a nonrotating star and a star rotating at 200~Hz is much smaller than the measurement uncertainty, for a wide range of gravitational masses and EOSs.  

If we wish to compare a measured mass with the maximum stable gravitational mass of a nonrotating star, then we need only to specify the EOS.  In contrast, if we need to compute the likelihood of a NICER mass-radius posterior with the expectations from a given EOS, then we must also assume a probability distribution for the central density of the star, given that this is the boundary condition for the TOV and related equations.  Here we follow the convention in \citet{2021ApJ...918L..28M}, in which for a given EOS the prior on the central density is quadratic between the central density $\rho_{\rm min}$ of a $1~M_\odot$ star and the maximum central density $\rho_{\rm max}$ of a stable star.  That is, if $x$ is distributed uniformly between 0 and 1, then the central density prior is given by $\rho_c=\rho_{\rm min}+x^2(\rho_{\rm max}-\rho_{\rm min})$, except that we skip densities that yield unstable stars (this can happen for $\rho_c<\rho_{\rm max}$ in, for example, EOS with phase transitions, which are included in our sample).  This contrasts with the linear central density prior chosen in \citet{2019ApJ...887L..24M}, but the two density priors lead in practice to very similar posteriors.

When we compute the likelihood of a NICER mass-radius posterior, or of a gravitational wave measurement of the tidal deformability $\Lambda$ as a function of neutron star mass, we need to integrate over the central densities.  For each central density we thus need to determine the probability of the observation given an EOS model.  Because we have discrete samples of posterior points in the $M-R$ or $M-\Lambda$ planes, it is necessary to employ a procedure that uses these discrete samples to estimate a smoothed probability distribution in two dimensions.  As in \citet{2019ApJ...887L..24M} and \citet{2021ApJ...918L..28M}, we use Gaussian kernel density estimation for this purpose, but with 0.1 times the standard bandwidth recommended by \citet{silverman1986}.  See Section~5.1 of \citet{2021ApJ...918L..28M} for additional details.

We also include a constraint of $S=32\pm 2$~MeV \citep{2012PhRvC..86a5803T} on the nuclear symmetry energy $S$ (defined as the difference in the total energy per nucleon between matter with an equal density of protons and of neutrons, and the total energy per nucleon of pure neutron matter) at nuclear saturation density.  \citet{2021ApJ...918L..28M} approximated $S$ as $S=\epsilon/n-m_nc^2$ at a number density $n$, which is equivalent to assuming that the matter at nuclear saturation density is pure neutrons.  Here we compute $S$ at saturation density by imposing beta equilibrium on the matter, including the possible presence of muons, using the formulae in Section~II of \citet{2016arXiv160408575B}.  This makes a few percent difference in $S$ but, as noted in \citet{2021ApJ...918L..28M}, the inclusion of a constraint on $S$ has only a minor effect on the EOS posteriors because it applies at comparatively low densities.

Some studies of the equation of state of high-density matter apply additional theoretical constraints, based, e.g., on chiral effective field theory (see \citealt{2021ARNPS..71..403D} for a recent review) or perturbative quantum chromodynamics (pQCD; see \citealt{2023ApJ...950..107G} for a recent example, but also see \citealt{2024EPJWC.29603002M,2024LRR....27....3K,2024PhRvD.110l3009M}).   Here we do not include such considerations, so that our samples can be used with different assumptions about nuclear physics. 

\subsection{EOS Models}
\label{sec:eosmodels}

We follow \citet{2024ApJ...974..295D} in using Gaussian processes to extend our EOS beyond half of nuclear saturation density (although we note that other choices are possible, such as piecewise polytropes or spectral forms; see \citealt{2019ApJ...887L..24M,2021ApJ...918L..28M}).  Gaussian processes were introduced in EOS estimates by \citet{2019PhRvD..99h4049L} and developed extensively and mostly by the same authors (e.g., \citealt{2020PhRvD.101f3007E,2020PhRvC.102e5803E,2020PhRvD.101l3007L,2021PhRvD.104f3003L,2023PhRvD.108d3013E,2024PhRvD.109b3020L,2025PhRvD.112j3023F}).  See \citet{2006gpml.book.....R} for an overall introduction to Gaussian processes.  The basic idea is that, starting with a fiducial function $f_0({\bf x})$ tabulated at particular values ${\bf x}$ of the independent variable, one can create variations $f({\bf x})=f_0+\Delta f({\bf x})$ on that function by drawing $\Delta f({\bf x})$ from a multivariate Gaussian distribution.  In the context of EOS estimation, following \citet{2010PhRvD..82j3011L} and \citet{2019PhRvD..99h4049L}, our independent variable is the log pressure and the dependent variable is 
\begin{equation}
\phi\equiv\ln\left(c^2\frac{d\epsilon}{dP}-1\right)\; ,
\end{equation}
which is constructed so that (as is necessary for a full Gaussian distribution) physically possible values of $\phi$ can range from $-\infty$ (corresponding to the boundary of causality, with a sound speed $c_s=(dP/d\epsilon)^{1/2}=c$) to $+\infty$ (corresponding to the boundary of stability $c_s=0$).  For consistency with \citet{2021ApJ...918L..28M} we choose as our fiducial function $\phi_0=5.5-2.0(\log_{10} P-32.7)$ (where the pressure $P$ is in cgs units).  This matches well the EOS listed at the CompOSE website https://compose.obspm.fr/table/families/3/ but also drives the sound speed to values approaching the speed of light at several times nuclear saturation density.  More flexible choices are possible; see for example \citet{2020PhRvD.101f3007E}.  We also need to choose a kernel which correlates the deviations $\Delta\phi_i$, $\Delta\phi_j$ from $\phi_0$ at two values $x_i$, $x_j$ of the independent variable $x=\ln P$.  That is, if the deviation from the fiducial value is drawn from
\begin{equation}
{\cal N}(0,\Sigma(\Delta \phi_i,\Delta \phi_j))\; ,
\end{equation}
where ${\cal N}$ represents a normal distribution, then we need to choose the covariance matrix $\Sigma$.  Following \citet{2021ApJ...918L..28M}, we assume that we can write $\Sigma(\Delta \phi_i, \Delta \phi_j)=K(x_i,x_j)$ and we choose a ``squared exponential" kernel
\begin{equation}
K(x,x^\prime)=\sigma^2\exp\left(-\frac{(x-x^\prime)^2}{2\ell^2}\right)
\end{equation}
with $\sigma=1$ and $\ell=1$.  We note finally that the flexibility of the Gaussian processes framework allows the selection of EOS with $dP/d\epsilon\ll c^2$ in finite density intervals (i.e., effectively, phase transitions; this is the approach taken by \citealt{2023PhRvD.108d3013E}) or the addition of sharp features in the dependence of sound speed on density that simulate phase transitions and other features \citep{2023JPhCS2536a2006M,2024PhRvD.110l3009M}.

\subsection{EOS results}
\label{sec:results}

\begin{figure}
          \includegraphics[width=\linewidth]{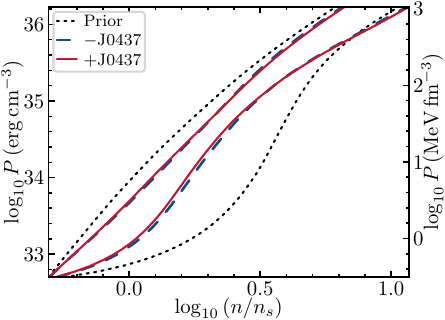}
\caption{Effect of our mass and radius measurement of PSR~J0437$-$4715 on the EOS of high-density matter, using the Gaussian process framework described in Section~\ref{sec:eosmodels}.  The black dotted lines show the prior on the EOS, the blue dashed lines show the EOS constraints without our analysis of the NICER data on PSR~J0437$-$4715 (using the measurements discussed in Section~\ref{sec:statistical}), and the red solid lines show the EOS constraints including our analysis of the PSR~J0437$-$4715 data.  Here we plot the log of the pressure (using different unit systems on the left-hand and on the right-hand axes) versus the log of the baryonic number density in units of the saturation density $n_s=0.16$~fm$^{-3}$.  For each line type, at a given density the lower curve gives the 5th percentile of the pressure and the upper curve gives the 95th percentile of the pressure.  We see that the inclusion of the PSR~J0437$-$4715 measurement tightens the EOS slightly in the $n\approx (1-3)n_s$ density range.  See the main text for additional details.}     
\label{fig:EOS}
\end{figure}

Figure~\ref{fig:EOS} summarizes the effect that our measurement of the mass and radius of PSR~J0437$-$4715 has on our knowledge of the EOS of high-density matter, using the Gaussian process framework described in Section~\ref{sec:eosmodels}.  In this figure we compare the 5th and 95th percentiles of the pressure as a function of baryon number density for the prior (black dotted lines), the prior updated with pre-J0437 measurements (blue dashed line; see Section~\ref{sec:statistical} for a discussion of the data that we include), and finally the prior additionally updated with the PSR~J0437$-$4715 measurements of the current paper (red solid lines).  We see that the addition of our PSR~J0437$-$4715 measurement tightens the EOS posterior from $n\approx (1-3)n_s$, but only slightly.  

\begin{deluxetable}{ccccc}
\setlength{\tabcolsep}{2.3pt}
\caption{Maximum Mass and Fiducial Radius including PSR~J0437$-$4715 measurement}
\tablehead{
\colhead{Quantity} & \colhead{Data set} &  \colhead{$-1\sigma$} & \colhead{median} & \colhead{$+1\sigma$} 
}
\startdata
$M_{\rm TOV}(M_\odot)$&\citet{2024ApJ...974..295D}&2.08&2.22&2.44 \\
$M_{\rm TOV}(M_\odot)$&This work&2.02 &2.18 &2.42 \\
$M_{\rm TOV}(M_\odot)$&PSR~J0952$-$0607&2.28 &2.42 &2.62 \\
\hline
$R_e(1.4~M_\odot)({\rm km})$&\citet{2024ApJ...974..295D}&12.09 &12.57 &13.06 \\
$R_e(1.4~M_\odot)({\rm km})$&This work&12.08 &12.56 &13.03 \\
$R_e(1.4~M_\odot)({\rm km})$&PSR~J0952$-$0607&12.08 &12.52 &12.99 \\
\enddata
\tablecomments{Updated $-1\sigma$, median, and $+1\sigma$ points in the posterior distributions of the maximum gravitational mass of a nonrotating neutron star ($M_{\rm TOV}$) and the equatorial circumferential radius of a fiducial nonrotating $1.4~M_\odot$ neutron star ($R_e$), compared with those inferred by \citet{2024ApJ...974..295D}.  In addition to our measurement of PSR~J0437$-$4715 from NICER data, the other update in the ``This work" rows is that we use the mass estimate for PSR~J0348$+$0432 of $M=1.806\pm 0.037~M_\odot$ from \citet{2025ApJ...983L..20S}, rather than the older estimate of $M=2.01\pm 0.04~M_\odot$ from \citet{2013Sci...340..448A} that was used in earlier papers.  We see that the combination of these two updates shifts both $M_{\rm TOV}$ and $R_e(1.4~M_\odot)$ to slightly smaller values.  For comparison, we also indicate the effect on the $M_{\rm TOV}$ and $R_e(1.4~M_\odot)$ posteriors if we include the updated mass constraint $M=2.35\pm 0.11~M_\odot$ on the black widow pulsar PSR~J0952$-$0607 from \citet{2025arXiv251205099R} (see text for details).  We see that inclusion of this mass would increase the estimated $M_{\rm TOV}$ substantially but would have very little effect on $R_e(1.4~M_\odot)$.}
\label{tab:maxmass}
\end{deluxetable}

Table~\ref{tab:maxmass} shows the effect of our updated measurement, as well as the update on the estimated mass of PSR~J0348$+$0432 (from $M=2.01\pm 0.04~M_\odot$ in \citealt{2013Sci...340..448A} to $M=1.806\pm 0.037~M_\odot$ in \citealt{2025ApJ...983L..20S}), on the estimate of the maximum mass of a nonrotating neutron star ($M_{\rm TOV}$) and on the radius of a fiducial $M=1.4~M_\odot$ neutron star ($R_e(1.4~M_\odot)$).  We see that the effect is small but that both $M_{\rm TOV}$ and $R_e(1.4~M_\odot)$ are shifted to slightly smaller values.

For comparison, in Table~\ref{tab:maxmass} we also show the effect on $M_{\rm TOV}$ and $R_e(1.4~M_\odot)$ if we include the updated mass $M=2.35\pm 0.11~M_\odot$ of the black widow pulsar PSR~J0952$-$0607, from \citet{2025arXiv251205099R}.  A nuance is that because PSR~J0952$-$0607 rotates very rapidly (707.31~Hz), there is extra rotational support.  We follow \citet{2025arXiv251205099R} in applying a correction of $0.03~M_\odot$ downward when constraining $M_{\rm TOV}$ (based on the relations presented by \citealt{2022ApJ...934..139K}), so we effectively use a nonrotating mass of $2.32\pm 0.11~M_\odot$.  We see that inclusion of this mass would increase the estimated $M_{\rm TOV}$ substantially, but would have little effect on $R_e(1.4~M_\odot)$.  We also note that \citet{2025arXiv251205099R} urge caution in using the masses of black widows, given possible systematics, but it is clear that if the masses can be considered reliable then they will provide important input.

\section{CONCLUSIONS}
\label{sec:conclusions}

Models of the NICER data on PSR~J0437$-$4715 require a modulated nonthermal component as well as  somewhat complicated spot configurations for the thermal X-ray emission.  This complexity offsets the superb data from NICER and the excellent radio measurements from this binary (which provide tight constraints on the mass, the distance, and the observer inclination) to yield a relatively unconstraining radius posterior.  This posterior is, however, fully consistent with NICER measurements of the pulsar PSR~J0030$+$0451, which has a similar mass of $M\sim 1.4~M_\odot$.  These measurements, combined with previous analyses of NICER data, the constraints on the tidal deformability of the neutron stars in the gravitational wave event GW170817, and the high masses inferred for several pulsars, have dramatically improved our understanding of cold, catalyzed matter beyond nuclear saturation density.  We anticipate that work in progress on data-efficient filtering and background modeling, together with additional data collection, is likely to result in a future modest but significant improvement in NICER's radius constraint for PSR~J0437$-$4715.

\section*{Acknowledgments}

A.J.D. and M.C.M. were supported in part by NASA ADAP grant No. 80NSSC21K0649.  Part of this work was performed at the Aspen Center for Physics, which is supported by National Science Foundation grant PHY-2210452.  We also thank the Institute for Nuclear Theory at the University of Washington for its kind hospitality and stimulating research environment. This research was supported in part by the INT's U.S. Department of Energy grant No.~DE-FG02-00ER41132. A.J.D. was supported by NASA through the Hubble Fellowship Program grant No.~HST-HF2-51553.001, awarded by the Space Telescope Science Institute, which is operated by the Association of Universities for Research in Astronomy, Inc., for NASA, under contract NAS5-26555. J.B. was supported by NASA under award number 80GSFC21M0002. S.B. was supported in part by NASA grants 80NSSC20K0275, 80NSSC22K0728, and 80NSSC25K7544.  W.C.G.H. acknowledges support through grant 80NSSC23K0078 from NASA. The research of S.M.M. is supported by NSERC Discovery Grant RGPIN-2019-06077. C.C. and Z.W. were supported by NASA under award number 80GSFC24M0006. Some of the resources used in this work were provided by the NASA High-End Computing (HEC) Program through the NASA Center for Climate Simulation (NCCS) at Goddard Space Flight Center.  The authors acknowledge the University of Maryland supercomputing resources (http://hpcc.umd.edu) that were made available for conducting the research reported in this paper. This research has made use of data products and software provided by the High Energy Astrophysics Science Archive Research Center (HEASARC), which is a service of the Astrophysics Science Division at NASA/GSFC and the High Energy Astrophysics Division of the Smithsonian Astrophysical Observatory.  We acknowledge extensive use of NASA's Astrophysics Data System (ADS) bibliographic services and arXiv.

\facility{\textit{NICER} (\citealt{2016SPIE.9905E..1HG})}

\software{emcee (\citealt{2013PASP..125..306F}), Python and NumPy (\citealt{2007CSE.....9c..10O}), Matplotlib (\citealt{2007CSE.....9...90H}), PGF/Ti$k$Z \citep{tantau:2013a}, Cython (\citealt{2011CSE....13b..31B}), schwimmbad (\citealt{schwimmbad}), HEASoft \citep{2014ascl.soft08004N}, PINT \citep{2021ApJ...911...45L}, and pocoMC (\citealt{2022MNRAS.516.1644K})}

\pagebreak

\appendix

\section{Breakdowns of fits by components}

\begin{figure}
          \includegraphics[width=1.0\linewidth]{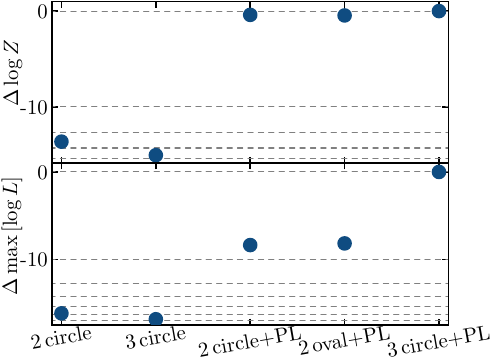}
\caption{Natural log of evidence (top panel) and natural log of maximum likelihood (bottom panel) for different models, relative to our featured model (three uniform-temperature circular spots plus a modulated Gaussian-profile power law).  From the left, these models are numbers 1, 2, 3, 4, and 5 in our list (see Sec.~\ref{sec:models} and Table~\ref{tab:model-list}).  Models 1 and 2, which do not include a modulated power-law component, are substantially disfavored compared with the models that do include this component.  Therefore, to allow those models to fit on the plots, the $-10$ to 0 portions of both panels are linear, whereas the portions below $-10$ are logarithmic (first horizontal dashed line is 10 below the reference model; second is 20 below; and so on).  We conclude that the modulated power law component must be included in a viable model of this source.  We also see that, in terms of the evidence, the two-circle, two-oval, and three-circle models (all with a modulated power law) are all comparable with each other.  The three-circle model, however, has a substantially larger maximum log likelihood than the other models.
}       
\label{fig:modelcompare}
\end{figure}

Figure~\ref{fig:modelcompare} compares the log evidence and maximum log likelihood for our models.  This comparison, like Table~\ref{tab:model-results}, demonstrates that, at least for our set of models, a modulated power law component dramatically improves the fit.

\begin{figure}
          \includegraphics[width=\linewidth]{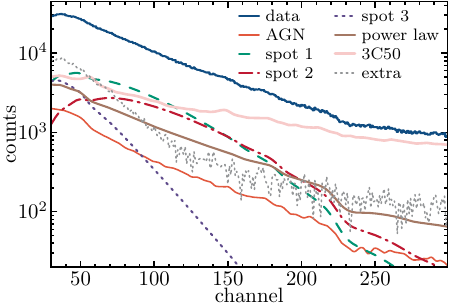}

\caption{Contribution of each fit component to the phase-summed count spectrum, as a function of NICER PI channel, using our best fit to the data of our model with three uniform-temperature circular spots plus a modulated power law, i.e., our model number 5.  The solid blue line shows the data, the fainter solid red line shows the contribution from the empirical NICER background of \citep{2022AJ....163..130R}, the brighter solid red line shows the contribution from the angularly nearby AGN, the dashed green line shows the contribution from the first spot, the dot-dashed red line shows the contribution from the second spot, the bold dotted black line shows the contribution from the third spot, the solid brown line shows the contribution from the modulated power law, and finally the light dotted line shows the remaining contribution, from the extra unmodulated background.  Note that the counts axis is logarithmic.  This plot demonstrates that multiple components have importance in different channel ranges: the extra (unidentified) unmodulated background is most prominent below channel $\sim 60$, the spots have their greatest relative importance at channels $\sim 60$--120, and although the 3C50 background is the most important single component above channel $\sim 120$, the remaining components all contribute (except for the low-temperature spot 3).  In particular, we see that the power law is the most important of the modulated components above channel $\sim 200$.  The apparent erratic nature of the extra component is due to Poisson fluctuations in the data, which are enhanced when other components are subtracted.
}     
\label{fig:spectrum}
\end{figure}

\begin{figure}
          \includegraphics[width=\linewidth]{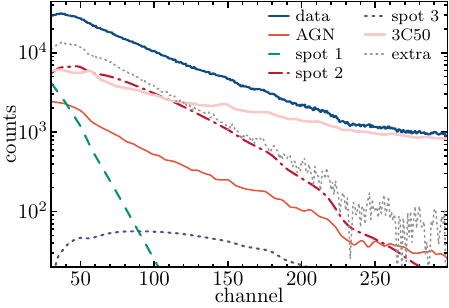}

\caption{The same as Figure~\ref{fig:spectrum}, but for the best three-circle fit with no modulated power law.  Note that the spots have exchanged roles, e.g., Spot 1 is now the lowest-temperature spot.  The overall fit to the spectrum is as good as the fit with a model including a modulated power law, because of the flexibility that we assume in the extra unmodulated background.  However, because of the lack of the modulated power law component, the extra background has more counts.}
  
\label{fig:spectrumnoPL}
\end{figure}

Figure~\ref{fig:spectrum} shows the contribution of each of the fit components to the overall best fit, as a function of NICER PI channel.  The upper blue solid line shows the data; the best-fit model, which is the sum of the components, is very close to the data.  The three spots are represented by the green dashed line, the red dot-dashed line, and the blue dotted line.  The known background components are shown with the red solid line (from the angularly nearby AGN) and the pink solid line (the empirical background of \citealt{2022AJ....163..130R}).  The brown solid line shows the contribution of the modulated power law, and the gray dotted line shows the additional unmodulated background that we add as part of the fit (note that the vertical axis is logarithmic, and that the extra component has picked up the Poisson fluctuations in the data).  All of the components are important in at least part of the channel range.  Figure~\ref{fig:spectrumnoPL} shows the same plot for the best three-circle model with no modulated power law, where we see that the required extra background is larger than it was when the modulated power law was included.

\begin{figure}
          \includegraphics[width=\linewidth]{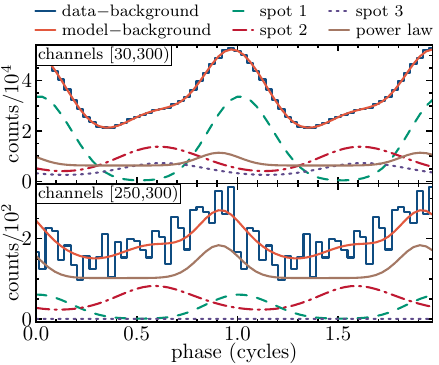}

\caption{Waveform after subtraction of the unmodulated background components (from the 3C50 background, the angularly nearby AGN, and the ``extra" component in Figure~\ref{fig:spectrum}), using our best fit to the data of our model with three uniform-temperature circular spots plus a modulated power law.  The upper panel shows the full bolometric waveform, whereas the lower panel shows the waveform summing NICER PI channels 250 through 299 inclusive.  The modulated power law is most important in the higher-energy channels, but it also contributes non-negligibly to the full bolometric waveform.}     
\label{fig:waveform}
\end{figure}

\begin{figure}
          \includegraphics[width=\linewidth]{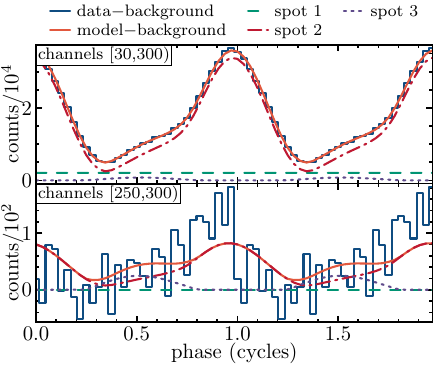}

\caption{Same as Figure~\ref{fig:waveform}, but for the best three-circle fit that does not include a modulated power law component.  Because the spectral contribution from the modulated power law does not exist in this fit, the total background must account for these counts and thus the data minus background and model minus background count totals are lower than they were in Figure~\ref{fig:waveform}.  We also see that, without the modulated power law, the fit in the higher-energy channels is considerably worse than it was in Figure~\ref{fig:waveform}.}     
\label{fig:waveformnoPL}
\end{figure}

Figure~\ref{fig:waveform} shows the bolometric (top panel) and high-energy (lower panel) waveform for the data, the full model, and the model components, after removal of the known and added unmodulated background contributions.  This leaves the three spots and the modulated power law.  We see that each component contributes; the modulated power law is most important at higher energies, but also cannot be neglected in the bolometric waveform.  Figure~\ref{fig:waveformnoPL} shows the same comparison, but for the best three-circle model without a modulated power law.

\section{Posterior Distributions}
\label{sec:appendix}

\begin{figure*}
     \includegraphics[width=\linewidth]{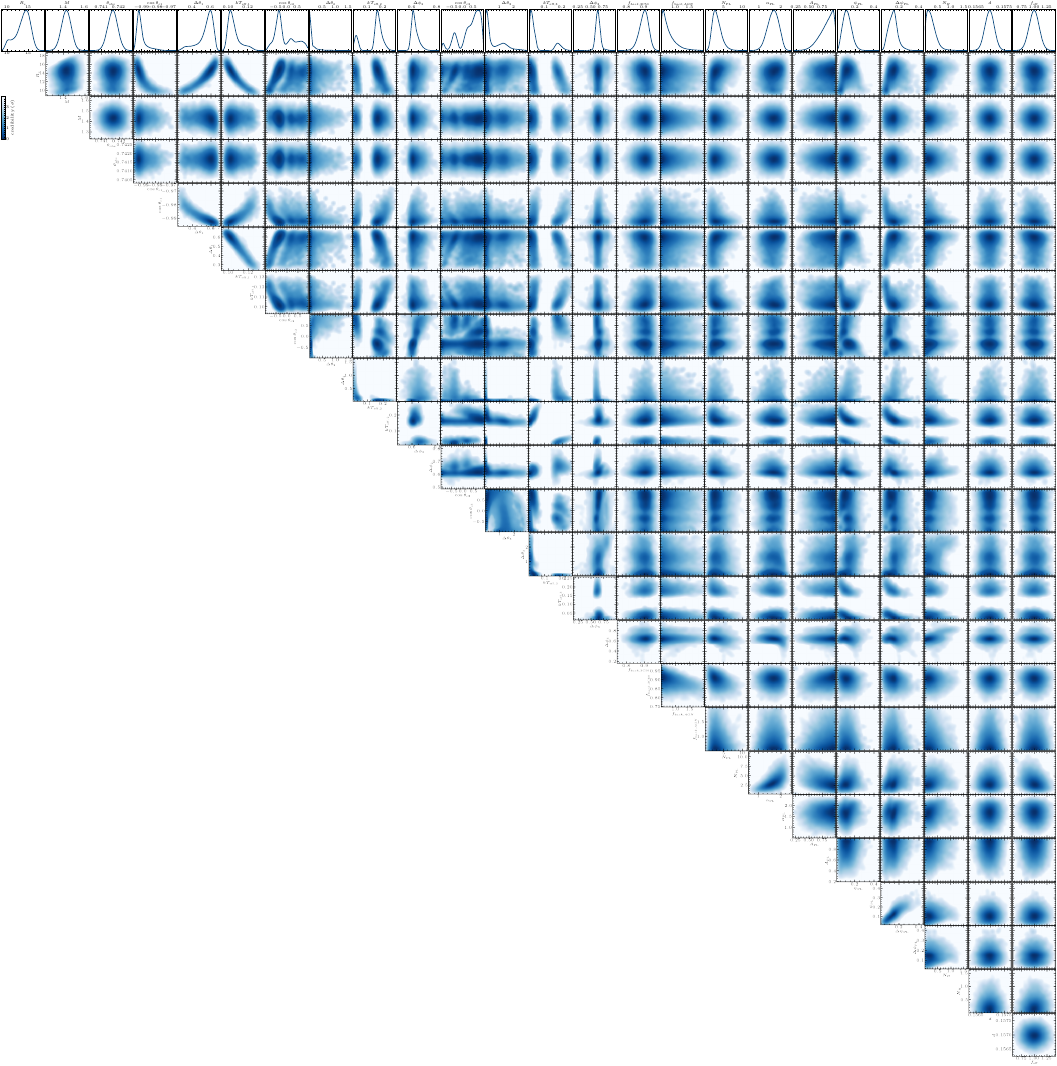}
    \caption{Posterior probability density distributions from our analysis of the NICER data on PSR~J0437$-$4715.  Here we use our featured model, which has three uniform-temperature circular hot spots plus a modulated power-law spectrum with a Gaussian profile in rotational phase.  The power-law normalization $N_{\rm PL} $ is in units of $10^{-5}~{\rm keV}^{-1}~{\rm cm}^{-2}~{\rm s}^{-1}$.}
    \label{fig:posteriors}    
\end{figure*}

Table~\ref{tab:posteriors} lists the median, $\pm1\sigma$, and $\pm2\sigma$ points in the posterior distributions obtained by fitting our models to only the NICER data, assuming a fully ionized hydrogen atmosphere, and the resulting maximum likelihood values for each of the parameters in these models. We display the complete corner plot of the posteriors from these same analyses in Figure~\ref{fig:posteriors}.

\renewcommand{\arraystretch}{0.875}
\setlength{\tabcolsep}{2.pt}
\begin{deluxetable}{ccccccc}
\tablecaption{Fits to NICER Data on PSR~J0437$-$4715}
\tablehead{
\colhead{Parameter} & \colhead{median} & \colhead{$-1\sigma$} & \colhead{$+1\sigma$} & \colhead{$-2\sigma$} & \colhead{$+2\sigma$} & \colhead{Maximum Likelihood} 
}
\startdata
$R_e\,\rm (km)$& 13.892 & 11.823 & 15.142 & 9.927 & 16.032 & 15.156 \\
$GM/c^2R_e$& 0.152 & 0.139 & 0.179 & 0.132 & 0.213 &  0.143\\
$M\,(M_\odot)$& 1.431 & 1.388 & 1.474 & 1.345 & 1. 517 & 1.470 \\
$\cos\theta_{\rm c1}$ &$-0.9914$ & $-0.9928$ & $-0.9880$ & $-0.9938$ & $-0.9800$ & -0.9928 \\
$\Delta\theta_1\,\rm (rad)$&0.571 & 0.471 & 0.615 & 0.317 & 0.639 & 0.610 \\
$kT_{\rm eff,1}\,\rm (keV)$&0.105 & 0.101 & 0.112 & 0.099 & 0.123 & 0.106 \\
$\cos\theta_{\rm c2}$&$-0.277$ & $-0.499$ & $0.538$ & $-0.796$ & $0.886$ & 0.847 \\
$\Delta\theta_2\,\rm (rad)$&0.012 & 0.008 & 0.127 & 0.006 & 0.578 & 0.0097 \\
$kT_{\rm eff,2}\,\rm (keV)$&0.167 & 0.039 & 0.192 & 0.020 & 0.223 & 0.172 \\
$\Delta\phi_2\,\rm (cycles)$&0.619 & 0.604 & 0.652 & 0.592 & 0.693 & 0.638 \\
$\cos\theta_{\rm c3}$&$0.296$ & $-0.501$ & $0.781$ & $-958$ & $0.958$ & 0.925 \\
$\Delta\theta_3\,\rm (rad)$&0.208 & 0.012 & 1.283 & 0.007 & 2.195 & 0.139 \\
$kT_{\rm eff,3}\,\rm (keV)$&0.031 & 0.018 & 0.171 & 0.016 & 0.209 & 0.043 \\
$\Delta\phi_3\,\rm (cycles)$&0.647 & 0.613 & 0.684 & 0.582 & 0.762 & 0.656 \\
$f_{\rm back,3C50}$ & 0.903 & 0.876 & 0.926 & 0.845 & 0.947 &  0.828\\
$f_{\rm back,AGN}$ & 0.681 & 0.545 & 0.977 & 0.506 & 1.439 & 0.566 \\
$N_{\rm PL} (10^{-5}~{\rm keV}^{-1}~{\rm cm}^{-2}~{\rm s}^{-1})$& 2.964 & 2.062 & 4.213 & 1.424 & 5.851 & 6.373 \\
$\alpha_{\rm PL}$ & 1.649 & 1.389 & 1.893 & 1.127 & 2.086 & 2.156 \\
$A_{\rm PL}$ & 0.859 & 0.693 & 0.960 & 0.518 & 0.994 & 0.803 \\
$\psi_{\rm PL}$ & 0.210 & 0.093 & 0.177 & 0.062 & 0.244 & 0.100 \\
$\Delta\psi_{\rm PL}$ & 0.138 & 0.093 & 0.177 & 0.062 & 0.244 & 0.102 \\
$\theta_{\rm obs}\,\rm (rad)$& 0.7417 & 0.7414 & 0.7419 & 0.7411 & 0.7422 & 0.7416 \\
$N_H\,(10^{20}\,\rm cm^{-2})$&0.211 & 0.070 & 0.407 & 0.011 & 0.676 & 0.563 \\
$d\,\rm (kpc)$&0.1570 & 0.1568 & 0.1571 & 0.1567 & 0.1573 & 0.1571 \\
$f_{\rm eff}$&1.000 & 0.902 & 1.098 & 0.805 & 1.195 & 0.825\\
\enddata
\label{tab:posteriors}
\tablecomments{A comparison of the $-2\sigma$, $-1\sigma$, median, $+1\sigma$, and $+2\sigma$, and maximum likelihood values inferred from our analysis of NICER data for PSR~J0437$-$4715.  Here we use our featured model, which has three uniform-temperature circular hot spots plus a modulated power-law spectrum with a Gaussian profile in rotational phase.}
\end{deluxetable}

\bibliographystyle{aasjournal}
\bibliography{references}

\end{document}